\documentclass[12pt]{article}
\usepackage{latexsym,graphicx,psfrag,here,cite}
\usepackage{bbm}
\usepackage{pifont} 
\usepackage{rotating}
\newcommand\lsim{\mathrel{\rlap{\lower4pt\hbox{\hskip1pt$\sim$}}
    \raise1pt\hbox{$<$}}}
\newcommand\gsim{\mathrel{\rlap{\lower4pt\hbox{\hskip1pt$\sim$}}
    \raise1pt\hbox{$>$}}}

\begin{document}

\begin{titlepage}
\begin{flushright} 
UWThPh-2007-2\\ 
\end{flushright}
\vskip 1.5cm

\begin{center} 
{\LARGE \bf Chiral Low-Energy Constants$^*$} \\[1cm] 

{\large  Gerhard Ecker} \\[.5cm] 

Faculty of Physics, Univ. of Vienna \\[.2cm]
Boltzmanngasse 5, A-1090 Vienna, Austria 
\end{center} 

\vfill

\begin{abstract}
\noindent 
The progress in determining the coupling constants of mesonic
chiral Lagrangians is reviewed, with emphasis on the work
performed in three successive European Networks (Eurodaphne I and II,
Euridice). Reliable estimates of those constants are essential for
making full use of next-to-next-to-leading-order calculations in 
chiral perturbation theory. The precision in the values of the strong
coupling constants of $O(p^4)$ has been increasing
steadily over the years. The situation is
less satisfactory in the nonleptonic weak sector where further
phenomenological input and more theoretical work are needed. A lot of
progress has recently been achieved for electromagnetic coupling constants
occurring in radiative corrections for mesonic processes at
low energies. 
\end{abstract}

\vfill
\noindent 
$^*$Presented at The Final Euridice Meeting, Kazimierz, Poland,
Aug. 2006 \\[.2cm] 
Supported in part by EU contract
No. MRTN-CT-2006-035482 (FLAVIAnet)
\end{titlepage}

\newpage

\section{Introduction}
Chiral low-energy constants (LECs) are the coupling constants of
effective chiral Lagrangians. 
They are independent of the light quark 
masses by construction and they describe the influence of all ``heavy'' 
degrees of freedom that are not contained explicitly in the effective  
Lagrangians. The construction of effective Lagrangians is based on
symmetry considerations only so that a lot of information is lost in 
going from the underlying Standard Model to the effective theory. As a 
consequence, effective Lagrangians contain many LECs, especially at
higher orders in the chiral expansion. Progress in chiral perturbation
theory (CHPT) \cite{Weinberg:1978kz,GLsu2,GLsu3}
depends on realistic estimates of chiral LECs.

My task in this talk was to review the progress in
determining or estimating chiral LECs since 1993 when the first
Eurodaphne Network got started. Most of this progress is in fact due
to work performed in the three European Networks Eurodaphne I, 
Eurodaphne II and Euridice. I will only consider the meson sector
here. The corresponding effective chiral Lagrangian is given in 
Table \ref{tab:Leff}. For a recent review including the baryon sector
see Ref.~\cite{Bernard:2006gx}. 

\begin{center} 
\renewcommand{\arraystretch}{1.2}
\begin{table}[H]
$$
\begin{tabular}{|l|c|} 
\hline
&  \\
\hspace{1cm} ${\cal L}_{\rm chiral\; order}$ 
~($\#$ of LECs)  &  loop  ~order \\[8pt] 
\hline 
&  \\
 ${\cal L}_{p^2}(2)$~+~ $
{\cal L}_{p^4}^{\rm odd}(0)$
~+~ $ {\cal L}_{G_Fp^2}^{\Delta S=1}(2)$  
~+~ ${\cal L}_{e^2p^0}^{\rm em}(1)$
~+~ ${\cal L}_{G_8e^2p^0}^{\rm emweak}(1)$ & $L=0$ \\[15pt]
~+~ ${\cal L}_{p^4}(10)$~+~
${\cal L}_{p^6}^{\rm odd}(23)$
~+~${\cal L}_{G_8p^4}^{\Delta S=1}(22)$
~+~${\cal L}_{G_{27}p^4}^{\Delta S=1}(28)$ &   
$L \le 1$ \\[3pt]
~+~ ${\cal L}_{e^2p^2}^{\rm em}(13)$
~+~${\cal L}_{G_8e^2p^2}^{\rm emweak}(14)$ 
~+~ ${\cal L}_{e^2p}^{\rm leptons}(5)$  & \\[15pt] 
~+~ ${\cal L}_{p^6}(90)$  & $L \le 2$ \\[8pt] 
\hline
\end{tabular}
$$
\caption{Effective chiral Lagrangian in the meson sector. The numbers
  in brackets refer to the number of LECs for chiral $SU(3)$.}
\label{tab:Leff}
\end{table}
\end{center}

Information on chiral LECs is obtained either from phenomenology or
with additional input from theory.
\begin{dinglist}{108}
\item Extraction from experiment \\[.1cm]
Some LECs are associated with terms in the Lagrangian that contribute
to amplitudes even in
the chiral limit. They govern the momentum dependence of amplitudes
and are at least in principle accessible experimentally. The
other class involves chiral symmetry breaking terms that specify the
quark mass dependence of amplitudes. The related LECs are much more
difficult to determine phenomenologically but they are accessible in
lattice simulations. 
\item Input from theory 

\vspace*{-.3cm}  

\begin{dinglist}{88}
\item Large-$N_c$ methods match CHPT with QCD by bridging
  the gap $M_K \lsim E \lsim 1.5$ GeV with resonance exchange. 
\item Lattice QCD.
\end{dinglist} 
\end{dinglist}

\section{Strong interactions}
At lowest order in the chiral expansion, there are only two LECs $B$
and $F$. $B$ depends on the QCD renormalization scale
and always appears multiplied by quark masses in CHPT amplitudes. The
products $B m_q$ and the constant $F$ can be expressed in terms of 
meson masses and of the pion decay constant $F_\pi$.
Those relations involve LECs of $O(p^4)$ or higher to be
discussed subsequently.

The strong chiral Lagrangians of $O(p^4)$ contain 7 measurable LECs
$l_i$ for chiral $SU(2)$ \cite{GLsu2} and 10 LECs $L_i$ for chiral
$SU(3)$ \cite{GLsu3}. The current
phenomenological values are based on calculations to $O(p^6)$
in most cases, sometimes supplemented by dispersive methods. Although
I restrict the discussion here to $SU(3)$, the most precise
determinations have been obtained for the $SU(2)$ LECs $l_1, l_2, l_4$ 
by combining CHPT to $O(p^6)$ with Roy equations \cite{CGL01}. This 
information can also be used for some of the $SU(3)$ LECs. The 
relations between
the $l_i$ and the $L_i$ are however only known \cite{GLsu3} to
$O(p^4)$, which is not sufficient for the present purpose.
Work in progress by the Bern group \cite{GHIS07} will soon 
provide those relations to $p^6$ accuracy.  

The present values of the renormalized $SU(3)$ LECs 
$L_i(M_\rho)$ are shown in the second column of Table
\ref{tab:LECSp4}. The first column contains the values originally
obtained in Ref.~\cite{GLsu3}. No drastic changes have occurred
although the mean values have generally decreased in absolute
magnitude. For $L_1, L_2, L_3$ the information from $\pi\pi$
scattering \cite{CGL01} will be very useful once the relations between
the $SU(3)$ and $SU(2)$ LECs will be available at the $p^6$ level 
\cite{GHIS07}. 

\begin{sidewaystable}[H]
\centering
\renewcommand{\arraystretch}{1.2}
\begin{tabular}{|c||rrrrcc|}  \hline 
 & & & & & &  \\[-6pt]
i & $O(p^4)$  \hspace*{.2cm} & $O(p^6)$ 
\hspace*{.3cm}   & 
$\pi K$  \hspace*{.5cm}& \mbox{} \hspace*{.9cm} lattice \hspace*{.6cm}
  & \mbox{} \hspace*{.1cm}  Ref.~\cite{Ecker:1988te} \hspace*{.1cm}
&  Ref.~\cite{Kaiser:2005eu} \hspace*{.0cm} 
\\[4pt] 
\hline
 & & & & & &  \\[-4pt]
  1  & 0.7 $\pm$ 0.3 & 0.43 $\pm$ 0.12 & 1.05 $\pm$ 0.12 &  &
     0.6  & 0.9    \\[2pt]
  2  & 1.3 $\pm$ 0.7 &  0.73 $\pm$ 0.12 & 1.32 $\pm$ 0.03 &  & 
     1.2   &  1.8 \\[2pt]
  3  & \hspace*{.5cm} $-$4.4 $\pm$ 2.5 & \hspace*{.7cm} $-$2.35 $\pm$
     0.37 & \hspace*{.4cm}  $-$4.53 $\pm$ 0.14  &  & 
     \hspace*{.3cm} $-$3.0 \hspace*{.6cm} & \hspace*{.3cm} $-$4.3 
     \hspace*{.55cm}  \\[2pt]
  4  & $-$0.3 $\pm$ 0.5 & $\sim$ 0.2 \hspace*{.4cm} & 0.53 $\pm$
     0.39  &  $-$0.2 $\pm$ 0.4 \hspace*{.2cm} & 0 \hspace*{.1cm} & 
     0 \hspace*{.1cm}   \\[2pt] 
  5  & 1.4 $\pm$ 0.5  &  0.97 $\pm$ 0.11 &  & 1.2 $\pm$ 0.4
     \hspace*{.2cm} &  
     1.4$^\ddagger$\hspace*{-.15cm}   &  2.2   \\[2pt]
  6  & $-$0.2 $\pm$ 0.3 & $\sim$ 0.0 \hspace*{.4cm}   & & 0.1 $\pm$
     0.2  \hspace*{.2cm} & 0 \hspace*{.1cm} & 0 \hspace*{.1cm}  \\[2pt]
  7  & $-$0.4 $\pm$ 0.2 &  $-$0.31 $\pm$ 0.14 &  & & 
     $-$0.3\hspace*{.25cm} &  $-$0.3 \hspace*{.15cm}  \\[2pt]
  8  & 0.9 $\pm$ 0.3 &  0.60 $\pm$ 0.18 &  & 0.7 $\pm$ 0.2
     \hspace*{.2cm} &  0.9$^\ddagger$\hspace*{-.15cm} &  0.8   \\[2pt]
  9   & 6.9 $\pm$ 0.7 & 5.93 $\pm$ 0.43 &  & & 
     6.9$^\ddagger$\hspace*{-.15cm} & 7.2  \\[2pt]
 10  & $-$5.5 $\pm$ 0.7 & $-$5.09 $\pm$ 0.47  &   & &
     $-$6.0 \hspace*{.1cm} & $-$5.4 \hspace*{.1cm}  \\[4pt]
\hline
\end{tabular}

\caption{Phenomenological values and theoretical estimates for the
$SU(3)$ LECs $L_i(M_\rho)$ in units of $10^{-3}$. The first column
  shows the original values of Ref.~\cite{GLsu3}, the second displays
  the present values taken from Ref.~\cite{Bijnens:2006zp} and
  references therein. The third column is based on an analysis of 
  $\pi K$ scattering \cite{Buettiker:2003pp} (only experimental errors
  are shown). The fourth column
  contains recent lattice results from the MILC Collaboration 
  \cite{Bernard:2006zp}. The fifth column shows the
  resonance saturation results of Ref.~\cite{Ecker:1988te} and 
  the last column reproduces a systematic estimate of resonance 
  contributions to lowest order in $1/N_c$ \cite{Kaiser:2005eu}. The 
  entries marked with $^\ddagger$ were
  taken as input in Ref.~\cite{Ecker:1988te}.}
\label{tab:LECSp4}

\end{sidewaystable}


\noindent 
The LECs  
$L_1, \dots , L_4$ have also been extracted from $\pi K$ scattering,
based on a dispersive analysis \cite{Buettiker:2003pp} 
applied to a CHPT calculation of $O(p^4)$. The
results are displayed in the third column in Table \ref{tab:LECSp4}
where only experimental errors are shown. 

With most of the chiral LECs of $O(p^4)$ reasonably well known, can we
understand the specific values with additional theory input?
Lattice QCD has come a long way to determine some of the constants
directly from QCD. The fourth column in Table \ref{tab:LECSp4} shows
the most recent results of the MILC Collaboration
\cite{Bernard:2006zp} with three dynamical light (staggered)
quarks. The agreement with the phenomenological values in column 2
is indeed ``staggering''. 

A different approach makes use of the properties
of QCD at large $N_c$ where amplitudes can be expressed in
terms of (stable) resonance exchange. To illustrate the main features
of this approach \cite{Ecker:1988te,Peris:1998nj}, let us consider 
elastic meson-meson scattering,
specializing to a channel with $s \leftrightarrow u$ symmetry
(e.g.: $\pi^+ \pi^0 \to \pi^+ \pi^0$). From axiomatic field theory
(Froissart theorem) we know that the scattering amplitude $A(\nu,t)$
satisfies a once-subtracted forward dispersion relation in
$\nu=(s-u)/2$: 
\begin{equation} 
A(\nu,t=0) = A(0,0) + \displaystyle\frac{\nu^2}{\pi}
\displaystyle\int_0^\infty d\nu^{\prime\,2}
\displaystyle\frac{Abs~A(\nu^\prime,0)}{\nu^{\prime\,2} \left(\nu^2 -
\nu^{\prime\,2} \right)}   ~.
\label{eq:disp}
\end{equation}

\noindent 
Exchange of a resonance $R$ generates the absorptive part
\begin{equation}
Abs~A(\nu,0) = \pi c_R M_R^4 \delta(\nu^2- M_R^4)
\end{equation}
where the constant $c_R$ is related to the partial decay width
$\Gamma(R \to \pi\pi)$ in this case. Therefore, Eq.~(\ref{eq:disp})
gives rise to
\begin{equation} 
A(\nu,0) = A(0,0) + \displaystyle\frac{c_R \nu^2}{\nu^2- M_R^4}~. 
\end{equation}
On the other hand, resonance exchange on the basis of a chiral
resonance Lagrangian produces an amplitude of the general form
\begin{equation} 
A_R (\nu,0) = \displaystyle\frac{P_R(\nu^2)}{\nu^2- M_R^4} ~,
\end{equation}
with a polynomial $P_R(\nu^2)$ satisfying the on-shell condition 
$P_R(M_R^4) = c_R M_R^4$. Decomposing the polynomial $P_R(\nu^2)$ as
\begin{equation}
P_R(\nu^2)=P_R(M_R^4)+\left(\nu^2- M_R^4
\right) \overline{P_R}(\nu^2) ~,
\end{equation}
the condition 
$A_R (\nu,0) = A(\nu,0)$ requires $\overline{P_R}(\nu^2)$
to be a constant,
\begin{equation} 
\overline{P_R}(\nu^2) = A(0,0) + c_R ~,
\end{equation}
which will not be the case for a general resonance Lagrangian. 
Therefore, the short-distance constraint embodied in the
once-subtracted dispersion relation (\ref{eq:disp}) demands that in 
general an appropriate polynomial $P_c(\nu^2)$ be added to
$A_R(\nu,0)$: 
\begin{equation}
A_R (\nu,0) = P_c(\nu^2) + \overline{P_R}(\nu^2) + 
  \displaystyle\frac{P_R(M_R^4)}{\nu^2- M_R^4} ~.
\end{equation}
The counterterm polynomial $P_c(\nu^2)$ is fixed by the short-distance 
constraint to satisfy
\begin{equation}
P_c(\nu^2) + \overline{P_R}(\nu^2) = A(0,0) + c_R ~,
\end{equation}
ensuring at the same time the correct
low-energy behaviour of the resonance exchange amplitude:
\begin{equation}
A_R (\nu,0) = A(\nu,0) = A(0,0) - \displaystyle\frac{c_R}{M_R^4} \nu^2
+ O(p^8) ~. 
\end{equation}
The coefficient of $\nu^2$ depends only on the mass and on the 
partial decay width of the resonance and it defines the resonance
contribution  to a certain combination of the $L_i$.

With the proper choice of resonance fields, such counterterm
polynomials are not needed at $O(p^4)$ for the exchange of  
$V(1^{--})$,  $A(1^{++})$,  $S(0^{++})$ and $P(0^{-+})$ mesons 
\cite{EGLPR}, but they are unavoidable for $T(2^{++})$ and  
$A(1^{+-})$ exchange \cite{Toublan:1995bk,EZ}.

The example of elastic meson meson scattering raises the
legitimate question why one should bother at all with resonance
Lagrangians? It may seem like an unnecessary detour to use the
couplings of a resonance Lagrangian that have to be corrected by
short-distance constraints after all. The alternative is to study
Green functions directly with a large-$N_c$ inspired ansatz in the
first place. The main advantages of a Lagrangian approach are 
first of all that chiral symmetry is automatically guaranteed for 
the generated Green
functions and amplitudes and there is no need to impose chiral Ward
identities. At least as important from a practical
point of view is the possibility to integrate out the resonances once
and for all in the generating functional of Green functions (always to 
leading order in $1/N_c$), thereby generating all LECs of a given
order. Of course, the short-distance analysis still remains to be done.

The fifth column in Table \ref{tab:LECSp4} shows the original
resonance estimates of Ref.~\cite{Ecker:1988te}. The last column
contains more recent systematic estimates of resonance contributions 
to lowest order in $1/N_c$ \cite{Kaiser:2005eu}. Remembering that the
renormalization scale is not fixed at leading order in $1/N_c$, the
agreement between the resonance exchange contributions and the
phenomenological values in Table \ref{tab:LECSp4} is more than
satisfactory. Attempts to include
corrections of next-to-leading order in $1/N_c$ have also been made
\cite{JJSC,Portoles:2006nr}. 

To improve the precision of LECs of $O(p^4)$, 
realistic estimates for some LECs of $O(p^6)$ are also
needed. Several results have already been obtained by members of the
Euridice Collaboration \cite{Moussallam:1997xx,Bijnens:2003rc,
Cirigliano:2004ue,JP}. 
All possible resonance contributions
of the standard variety $V(1^{--})$,  $A(1^{++})$,  $S(0^{++})$,
$P(0^{-+})$ have recently been presented in
Ref.~\cite{Cirigliano:2006hb}. The short-distance analysis
remains to be done in many cases of interest.  
  
The odd-intrinsic-parity Lagrangian of $O(p^4)$ is given by the
Wess-Zumino-Witten Lagrangian ${\cal L}_{p^4}^{\rm
  odd}(0)$ \cite{Wess:1971yu}
in Table \ref{tab:Leff}. After several conflicting results
in the literature there is now a consensus that the corresponding 
Lagrangian of  $O(p^6)$ has 23 LECs \cite{Bijnens:2001bb}. 
Only partial results
are available for the numerical values of those constants, but the
most promising approach is again based on a short-distance analysis
with or without chiral resonance Lagrangians
\cite{Moussallam:1997xx,Ruiz-Femenia:2003hm}.

\section{Nonleptonic weak interactions}
The chiral Lagrangian of lowest order, $O(G_F p^2)$, contains two LECs
$g_8,g_{27}$ to describe nonleptonic weak decays of kaons
\cite{cronin67}. Especially
the value of the octet coupling $g_8$ is very sensitive to 
chiral corrections \cite{Kambor:1991ah}. Isospin breaking
corrections are potentially
important for the 27-plet coupling constant $g_{27}$. The present
status is presented in Table \ref{tab:g827}. Although different
isospin breaking contributions are sizeable the overall
corrections are small for both LECs.

The LECs of $O(G_F p^4)$ (22 couplings $N_i$ in the octet and 28
couplings $D_i$ in the 27-plet Lagrangians)
are much less known than their strong counterparts at $O(p^4)$. 
The most recent phenomenological analysis of those combinations that 
occur in the dominant $K\to 2 \pi, 3 \pi$ decays can be found in
Ref.~\cite{Bijnens:2004ai}. Many more LECs appear in rare $K$ decays 
and a phenomenological update is definitely needed here.

\begin{center} 
\renewcommand{\arraystretch}{1.2}
\begin{table}[H]
\begin{tabular}{|c|ccc|}
\hline
&  & & \\[-.4cm] 
 &  \hspace*{.3cm} IC ~$O(G_F p^2)$ \hspace*{.3cm}  &  \hspace*{.3cm} 
IC ~$O(G_F p^4)$  \hspace*{.3cm} & \hspace*{.3cm}  IB ~$O(G_F p^4)$  
\hspace*{.3cm} \\
\hline 
& & & \\[-.2cm] 
\hspace*{.1cm} $g_8$ \hspace*{.1cm}
& $5.09 \pm 0.01$ & $3.67 \pm 0.14$ &  $3.65 \pm 0.14$ \\[.3cm] 
$g_{27}$ &  $0.294 \pm 0.001$ &  $0.297 \pm 0.014$ & $0.303 \pm 0.014$
\\[.2cm]
\hline 
\end{tabular}
\caption{Octet and 27-plet couplings $g_8,
  g_{27}$ at $O(G_F p^2)$ and at next-to-leading order,
  $O(G_F p^4)$, without
  (IC) and with (IB) isospin breaking \cite{Cirigliano:2003gt}.}
\label{tab:g827} 
\end{table}
\end{center}

Resonance saturation of weak LECs \cite{Ecker:1992de} suffers from the 
obvious drawbacks that the weak resonance couplings are unknown and that
short-distance constraints are missing. Nevertheless, resonance
saturation provides at least a possible parametrization of the
LECs. The most systematic approach is based on
factorization (valid to leading order in $1/N_c$)
\cite{Pallante:2001he,Cirigliano:2003gt}  but 
higher-order corrections in $1/N_c$ may well be sizeable.

If the situation of the LECs of $O(G_F p^4)$ is already
unsatisfactory, even less is known about higher orders. However, the
leading (double) chiral logs of $O(G_F p^6)$ are known
\cite{Buchler:2005jk}.

\section{Dynamical photons}
Radiative corrections at low energies involve the effective Lagrangians
in Table \ref{tab:Leff} with superscripts {\it em}, {\it emweak} or
{\it leptons} (semileptonic decays).

In the presence of photons as dynamical degrees of freedom the chiral
counting is different from the purely strong or nonleptonic weak
cases. The lowest-order Lagrangian for electromagnetic corrections
to strong processes is of $O(e^2 p^0)$ with a single LEC
\cite{Ecker:1988te}  that can be determined from the $\pi^+ - \pi^0$
mass difference. The next-to-leading-order Lagrangian of $O(e^2 p^2)$ 
with 13 LECs $K_i$ was constructed by Urech \cite{Urech:1994hd}. By
convoluting pure QCD $n$-point functions ($n \le 4$) with the photon
propagator, sum rule representations were derived for all the $K_i$ 
\cite{Ananthanarayan:2004qk}. Numerical estimates for the $K_i$ are 
obtained by saturating the sum rules with resonance exchange. Since
the LECs $K_i$ are difficult to determine phenomenologically, the
systematic work of Ref.~\cite{Ananthanarayan:2004qk} is especially
important for controlling radiative corrections to strong processes at
low energies.

The situation is much less favourable for electromagnetic corrections
to nonleptonic weak processes. The single LEC of lowest
order, $O(G_8 e^2 p^0)$, related to the electromagnetic penguin
contribution \cite{Bijnens:1983ye}, is reasonably well known. However,
the 14
additional LECs of $O(G_8 e^2 p^2)$ \cite{Ecker:2000zr} are only known
to leading order in $1/N_c$ (factorization). In this way, the LECs can
be expressed in terms of Wilson coefficients, the strong LECs $L_5,L_8$
and the electromagnetic LECs $K_i$
\cite{Pallante:2001he,Cirigliano:2003gt}.

\section{Dynamical photons and leptons}
Radiative corrections for semileptonic weak decays require the
incorporation of leptons as dynamical degrees of freedom. The
leading-order Lagrangian ${\cal L}_{e^2p}^{\rm leptons}(5)$ in Table 
\ref{tab:Leff} contains five LECs $X_i$ \cite{Knecht:1999ag}.

With a two-step matching procedure (Standard Model 
~$\leftrightarrow$~  Fermi theory   ~$\leftrightarrow$~
CHPT), Descotes-Genon and Moussallam have recently established
integral representations for all the $X_i$
\cite{Descotes-Genon:2005pw}. 
One important application is in $K_{l3}$ decays, still the
best source for extracting the CKM matrix element $V_{us}$. As a
consistency check, the isospin violating ratio 
\begin{equation}
r_{+0} := \left( \frac{2 \, \Gamma(K^+_{e 3
(\gamma)}) \, M_{K^0}^5 \, I_{K^0}}{\Gamma(K^0_{e 3 (\gamma)}) \,
M_{K^+}^5 \, I_{K^+}} \right)^{1/2} =
\displaystyle\frac{|f_{+}^{K^+ \pi^0} (0)|}{|f_{+}^{K^0 \pi^-} (0)|} 
\end{equation}
has been considered \cite{Cirigliano:2004pv} that depends essentially
only on $X_1$. With the result for $X_1$ from
Ref.~\cite{Descotes-Genon:2005pw}, the theoretical
prediction $r_{+0} = 1.024 \pm 0.003$ is now in perfect agreement with
the most recent $K_{l3}$ data \cite{HN}. The agreement also indicates
that higher-order corrections to the theoretical
prediction for  $r_{+0}$ of $O[(m_u - m_d)p^4,e^2\,p^4]$ behave as
expected from chiral power counting.

\section{Outlook}
Since 1993, when Eurodaphne got started, substantial progress has
been made in the understanding of low-energy constants, both from
phenomenology and from theory (lattice QCD and
large-$N_c$ approaches).

In the strong sector, most of the machinery is now ready for a
precision determination of the LECs of $O(p^4)$. This endeavour
involves also LECs of $O(p^6)$ where we are still in the exploratory
stage. We need reliable estimates for some of those LECs to make full
use of 
next-to-next-to-leading-order calculations. In the nonleptonic sector,
improvements both in phenomenology and in theory are needed.
The most impressive progress in recent years has happened
for electromagnetic LECs. As a consequence, radiative corrections in
the meson sector at low energies are now under control. Semileptonic
$K_{l3}$ decays are one prime example of phenomenological relevance.

\noindent 
\section*{Acknowledgements}
\noindent 
I would like to pay a special tribute and to express my gratitude to 
Giulia Pancheri for having guided us successfully through three 
European Networks. Thanks are also due to Maria Krawczyk and Henryk 
Czy\.z for the efficient organization of the final Euridice Meeting.

\end{document}